\documentclass{JINST}

\usepackage{graphicx}

\title{Measurement and simulation of gamma-ray  background in a low energy accelerator facility }

\author{Sathi Sharma$^a$, M. Saha Sarkar$^a$\thanks{Corresponding author.}\\
\llap{$^a$}Saha  Institute  of  Nuclear  Physics, HBNI,
Kolkata - 700064, INDIA\\

  E-mail: \email{maitrayee.sahasarkar@saha.ac.in}}

\abstract{ 

The $\gamma$-ray background in the indoor environment  has been measured up to 3 MeV   
to evaluate the feasibility of studying  low cross-section (nanobarn to picobarn range) astrophysical reactions at
the Facility for Research in Experimental Nuclear Astrophysics (FRENA),  Saha Institute of Nuclear Physics, Kolkata.
An n-type coaxial HPGe detector with 20$\%$ relative efficiency has been placed at different locations in the accelerator and beam halls for the measurement. The measured activity has been compared with that at two laboratories (with standard brick walls) with and without passive and active radiation shieldings. As the halls at FRENA  are well shielded by concrete, the contribution of the shielding in indoor $\gamma$-ray background has been delineated by simulation using a 4$\pi$- geometry model. }


\begin{document}
\section{Introduction}
Natural $\gamma$-ray emitters contained in indoor and outdoor surrounding materials contribute significantly to the background of the $\gamma$-ray spectra.
Measurement and reduction of the background events are essential to improve the minimum detection level in the $\gamma$-ray spectrum.
Because of this, any $\gamma$-ray spectroscopy laboratory pays special attention to minimize the radiation background of detectors to improve 
the Minimum Detectable Activity (MDA) of the detection system. The MDA of the detection system is defined as \cite{Radulescu},
\begin{equation}
\label{eqn1}
 MDA \simeq \frac{\sqrt{Background~Counts}}{Detection~ Efficiency},
\end{equation}
Increasing the detector size increases efficiency to provide a lower limit in detectable activity (MDA), but it is not always an economically viable solution. Bigger detectors show increased summing effects. They are also more efficient in detecting background radiation. So, it is essential to optimize the parameters related to background suppression \cite{Britton} to minimize the MDA.

Omnipresent background radiation is emitted from a variety of natural and artificial radiation sources in the Earth, and its atmosphere and from cosmic rays originated from space. The energies of $\gamma$-rays coming from natural radioactive elements are generally up to 3 MeV. At energies above 3 MeV, the background in the $\gamma$-ray spectrum is primarily due to cosmic ray interactions within the upper atmosphere. 

Low background $\gamma$-ray spectroscopy is an essential tool for the study of rare nuclear physics processes. The meter water equivalent (often m.w.e.) is a standard unit to measure the effectiveness of shielding for cosmic ray attenuation in underground laboratories. Laboratories at the same depth of overburden can have different degrees of penetration of cosmic rays depending on their composition. The m.w.e. thus provides a convenient and consistent way of comparison of background suppression at different low background laboratories at overground and underground locations.

Low background $\gamma$-ray spectrometers are either operated in standard overground laboratories in buildings under a low overburden (<1m water equivalent (m.w.e.)),  or at specialized underground laboratories with high (>1000 m.w.e.) overburden. Even with a few meters water equivalent shielding, the soft component of cosmic rays can be shielded.

In nuclear astrophysics, all the relevant reactions take place far below the Coulomb barrier, and thus the reaction cross- sections are very small. The cross-sections are typically in the nanobarn - picobarn range \cite{caciolli}. So the reduction of events arising from background radiation is essential for these measurements. 
Thus the overburden of an underground laboratory may be necessary to shield the facility from cosmic radiation that can interfere with these experiments.
It is advantageous to work in an underground laboratory when the energies of $\gamma$-rays of interest are higher than  3 MeV. However,  if the $\gamma$-ray energies of interest are less than 3 MeV, then the underground facility does not provide any additional advantage \cite{caciolli}. Additional shieldings are needed to reduce the $\gamma$-rays originated from natural radioactivity.

Two kinds of radiation shieldings - designated as passive and active shielding,  are used for background $\gamma$-ray radiation suppression. In passive shielding, one generally covers the detector system with high Z elements, like lead (Pb, Z=82) with adequate thickness for the absorption of the highest energy environmental $\gamma$-ray radiation background. Layers of gradually  decreasing Z  cover the inner side of the shielding. 
The active shielding method includes sophisticated anti-correlation techniques in data acquisition. To stop the neutrons, which may give rise to subsidiary $\gamma$-ray radiation, low Z materials like paraffin, high-density polyethylene (HDPE), and concrete shielding are used. To optimize the shielding configuration, both the experimental technique and Monte Carlo modeling are essential.

A low energy accelerator-based Facility for Research in Experimental Nuclear Astrophysics (FRENA) is in the process of installation at the Saha Institute
of Nuclear Physics, Kolkata. The accelerator hall and beam hall in the building are shielded by a 1.2 m thick concrete wall to reduce the radiation dose outside these 
rooms to permissible limits. Nearly 60 cm ($\simeq$ 1.5 m.w.e.) of the concrete ceiling reduces the cosmic muons by a significant amount.
The energy loss of muons in concrete is nearly 4 MeV/cm \cite{Groom}, so muons with the energy of 240 MeV are entirely absorbed within the concrete shielding. However, we should keep in mind that these concrete walls work not only as  shields but are also sources of background 
$\gamma$-ray radiation originated from natural radioactivity. So, it is necessary to evaluate the extent of the contribution of these walls in the indoor environment of FRENA halls, especially in the $\gamma$-ray spectrum below 3 MeV.

Thus in the present work, we have measured the background $\gamma$-ray spectra at several positions of the FRENA accelerator building like accelerator hall and beam hall with a bare detector without any additional temporary shielding. Background spectra have been acquired at two other nuclear physics laboratories (Lab I and II) bounded by standard brick walls. The energy spectra have been compared to understand the shielding effect and the contributions from the concrete walls. Additionally, some passive and active shieldings have been arranged surrounding the HPGe detector. The suppression induced by these shieldings has been measured. 
The preliminary results have been reported in Ref. \cite{dae2018}. 

Monte Carlo simulation has been utilized to evaluate the contribution of these walls in the indoor environment of FRENA halls, especially in the $\gamma$-ray spectrum below 3 MeV. We have considered a spherical layer of concrete shielding as the simulation model geometry. The radioactive elements are isotropically distributed within the concrete walls. Similar 4$\pi$- geometry irradiation has been suggested 
for the radiation field in the buildings constructed with concrete walls by  Tsutsumi {\it et al.} \cite{Masahiro}. The effects of variations of different parameters like wall thickness, the radius of the spherical shell have been tested in the recent work.   Simulations have been done to optimize the shielding arrangements. 
  
\section{Sources of background radiations}
Background $\gamma$-ray radiation is emitted from a variety of natural and artificial radioactive sources. The primary natural sources are cosmic rays originated from space and sources in the Earth's crust, and its atmosphere. A tiny fraction of background $\gamma$-ray radiation originates from human-made sources.

The Earth's crust contains natural deposits of uranium, potassium, and thorium. The $^{238}$U,  $^{40}$K and $^{232}$Th nuclei are  radioactive and release ionizing radiation. Uranium, thorium, and potassium are omnipresent, {\it i.e.}, these nuclei are found essentially everywhere. Traces of these elements are also found in building materials.  The $\gamma$-ray spectrum up to 3 MeV is mainly dominated by the radiation emitted by $^{40}$K and the nuclei in the decay chains of $^{238}$U and $^{232}$Th etc. \cite{ndt}. Natural
potassium contains 0.012\% of $^{40}$K which decays with $T_{1/2} = 1.248 (3) \times 10^9$ y  \cite{ndt} and emits a $\gamma$-ray of 1.460 MeV energy \cite{ndt,Vishwanath}. The decay chain of  daughters of $^{238}$U ($T_{1/2}=4.468 (6) \times 10^9$ y) and $^{232}$Th ($T_{1/2} = 1.40 (1) \times 10^{10}$ y) emit
a long series of $\gamma$-rays \cite{ndt}. The highest energy intense $\gamma$-ray at energy 2.615 MeV, originates from the decay of  $^{208}$Tl,
a progeny of $^{232}$Th. One of the daughters of $^{238}$U,  $^{226}$Ra ($T_{1/2}$ = 1600 (7) y) has a long half-life which enhances the concentration of $^{222}$Rn and its progenies in the environment. $^{222}$Rn is an odorless and colorless radioactive gas. Radon (Rn) is an inert gas and thus does not react with surrounding matter. It can readily move up through the ground and accumulate in the ambient air of a closed room. Presence of $^{222}$Rn also gives rise to radiation background. 

The impact of cosmic rays in Earth's environment acts as an extraterrestrial source of high energy $\gamma$-ray background. Primary cosmic rays originate outside Earth's atmosphere. They contain about 99\%  simple protons (i.e., hydrogen nuclei) and alpha particles,  around  1\% of the nuclei of heavier elements and a tiny fraction of antimatter. Secondary cosmic rays are generated when primary cosmic rays interact with Earth's atmosphere. They consist of $\gamma$-rays and a large variety of elementary particles, including mesons, protons, neutrons, electrons, and positrons  \cite{wiki}. High energy $\gamma$-rays (E$>$ 3 MeV) and bremsstrahlung are produced when cosmic rays interact with the atmosphere of Earth and collide with ordinary matter.

In the present work,  we have measured the environmental $\gamma$-ray radiation background till 3 MeV.

\section{Experimental details}

\subsection{Venues of measurement}
\begin{figure}[ht]
\begin{center}
\includegraphics[width=\textwidth]{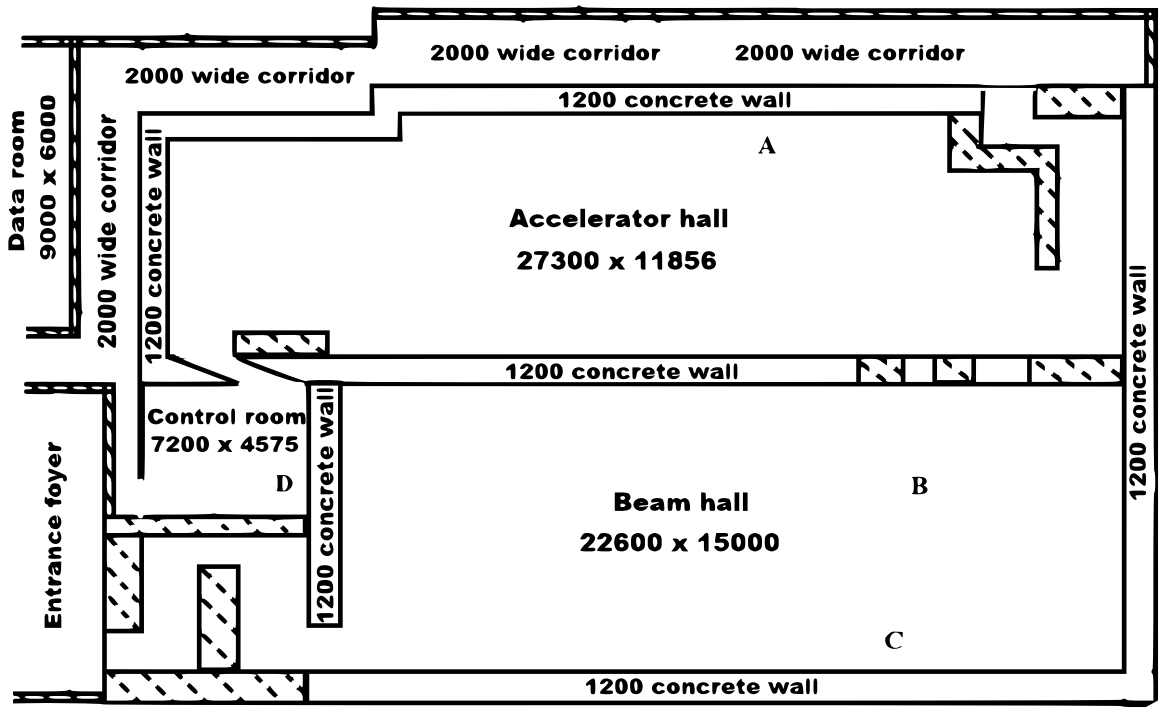}
\caption{\label{hall} \small \sl  The schematic diagram of the FRENA building ground floor.
A, B, C, and D are the locations of measurements. The boundaries with no hatched lines are permanent concrete shielding walls, and the hatched line ones are the temporary shielding walls.
All measurements are shown in mm. }
\end{center}
\end{figure}

The Facility for Research in Experimental Nuclear Astrophysics (FRENA) laboratory is a surface laboratory that is dedicated to 
low energy nuclear astrophysics experiments. It is a three storey building. It has four rooms: accelerator hall (27.300 m $\times$ 11.856 m), beam hall (22.600 m $\times$ 15.000 m), control room (7.200 m $\times$ 4.575 m), and data room (9.000 m $\times$ 6.000 m), respectively, at the ground floor. The associated laboratories are being developed on other floors. The accelerator hall and beam hall are well shielded with a 1.2 m thick concrete wall (Fig. \ref{hall}). 
The ground floor has a roof made of 60 cm thick concrete. 

The $\gamma$-ray radiation background has been measured at different locations (Fig. \ref{hall}),  in  the accelerator (A)  and  beam halls 
(B and C) as well as in the control room (C)  to get the integral background count rate at each place.

\begin{figure}[ht]
\begin{center}
\includegraphics[width=.5\linewidth]{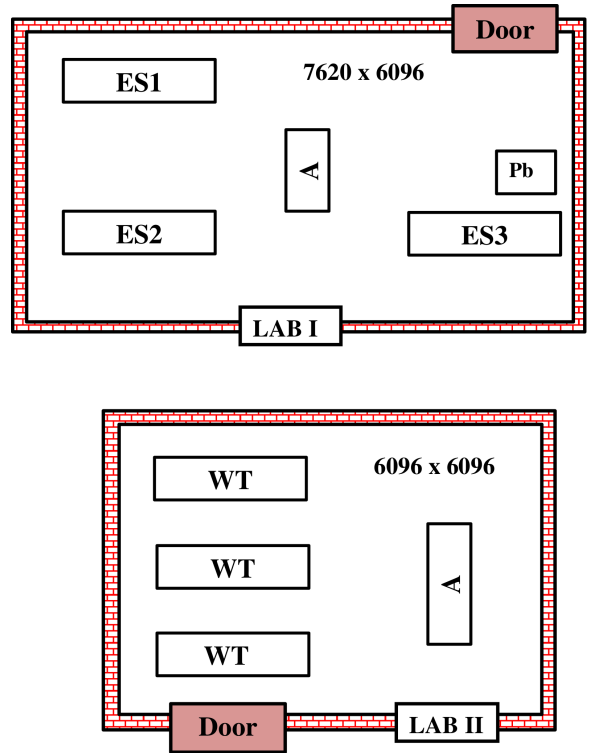}
\caption{\label{lab} \small \sl  The schematic diagram of the two laboratories, Lab I and Lab II, where A in each lab indicates the location of measurement. 
The working tables are indicated as WT, other experimental setups as ES1, ES2, and ES3. Pb indicates the Pb castle used for storing radioactive sources. }
\end{center}
\end{figure}

To compare the effect of concrete shielding in the accelerator and beam halls,  two nuclear physics laboratories (Lab I and II) with nearly 
25 cm thick standard brick walls (Fig. \ref{lab}) have been chosen in our present work. The rooms are situated on the first floor and second floor of a separate four - storey office building. The sizes of the laboratories are approximately one-third of the accelerator hall in the FRENA building. 

\subsection{The Detector and the  Data acquisition system}
An n-type coaxial HPGe detector with 80 cm$^{3}$ active volume and 20$\%$ relative efficiency \cite{relative_effi} was used for the measurement of $\gamma$-ray background at all the venues described earlier.  The relative efficiency of the HPGe detector, quoted by the manufacturer,  refers to the relative full energy photopeak (FEP) efficiency with respect to the absolute FEP  efficiency of a $3^" \times 3^" $ (7.62 cm $\times$ 7.62 cm) (diameter x height) NaI(Tl) crystal, for the 1.33 MeV peak of a $^{60}$Co source placed 25 cm from the detector. The HPGe detector crystal diameter is 5.1 cm, and the length is 4.2 cm. Experimental data have been acquired using a CAEN 5780M \cite{CAEN} desktop digitizer (14 bit, 16k channel, and 100 MS/s). The digitizer is equipped with DPP-PHA (Digital Pulse Processing-Pulse Height Analyzer) firmware that enables it to provide not only precise energy and timing information but also a portion of waveform and the other traces for fine-tuning of PHA settings \cite{CAEN}. 

\subsection{Experimental Setups}
\subsubsection{Setup at FRENA lab} 
At first, the HPGe detector has been placed at several locations in the FRENA accelerator building to check the effectiveness of the concrete wall and also its contribution in background $\gamma$-ray radiation. Data have been acquired with a bare HPGe detector without any additional passive or active shielding. The measurement time varied from several hours to days to have reliable counting statistics.
 
\subsubsection{Setups at Nuclear Physics laboratory}
Measurements of ambient  $\gamma$-ray radiation background have also been carried out at the nuclear physics laboratories (Lab I and Lab II). These laboratories are located at the second (Lab I) and first floor (Lab II) of a neighboring four storey building. 
At each location, data have acquired with a bare detector without and with passive and active shielding. The different setups are discussed below.
\begin{figure}[ht]
\begin{center}
\includegraphics[width =.5\linewidth,angle=90]{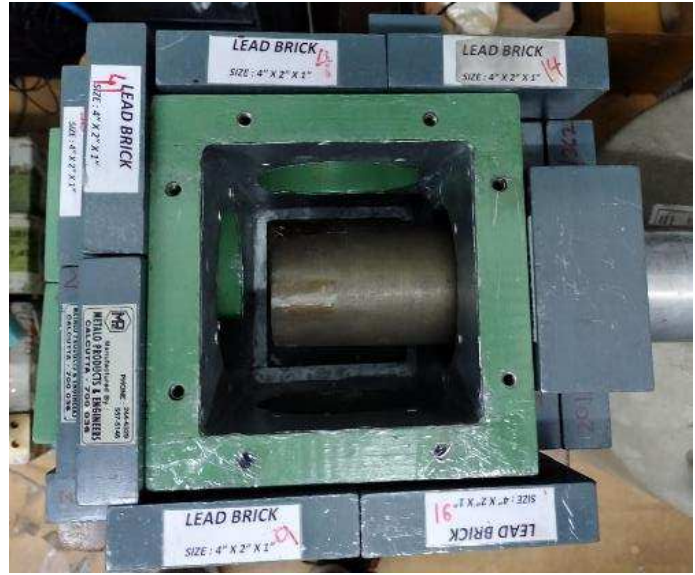}
\vspace{1cm}
\caption{\label{lab362} \small \sl  Top view (0.1 cm Cu liner and lead square box) of Setup at Lab I with passive shielding. }
\end{center}
\end{figure}

\begin{itemize}

\item{\it Setups at Lab I}
 \begin{itemize} 
\item{ No shielding:}\\
\noindent { Ambient  $\gamma$-ray radiation background has been acquired with the bare HPGe detector. The detector was placed away from the walls 
(marked as A in Lab I of Fig. \ref{lab}).
This laboratory has a few nuclear $\gamma$-ray spectroscopy setups (indicated as ES in Fig. \ref{lab}). The regularly-used laboratory radiation sources are also stored in this laboratory in a Pb castle (indicated as Pb in Fig. \ref{lab}). Thus the ambient background is higher than in the other neighboring rooms.}
\item{With passive shielding: }\\
\noindent {The detector is placed inside a 2 cm thick square box made of lead (Pb: Z=82). Additional 40 lead rectangular blocks, each with 2.5 cm thickness, have been used to cover the box from all sides. The effective thickness of Pb bricks is nearly 7 cm. The detector encasing 
is covered with 0.1 cm of Cu liner to cut the Pb X- rays. The inner-view  of the  shielding arrangement without the top cover is shown in 
Fig. \ref{lab362}}. 
\end{itemize}

\begin{figure}[ht]
\begin{center}
\includegraphics[width=10.0cm,height=10.0cm]{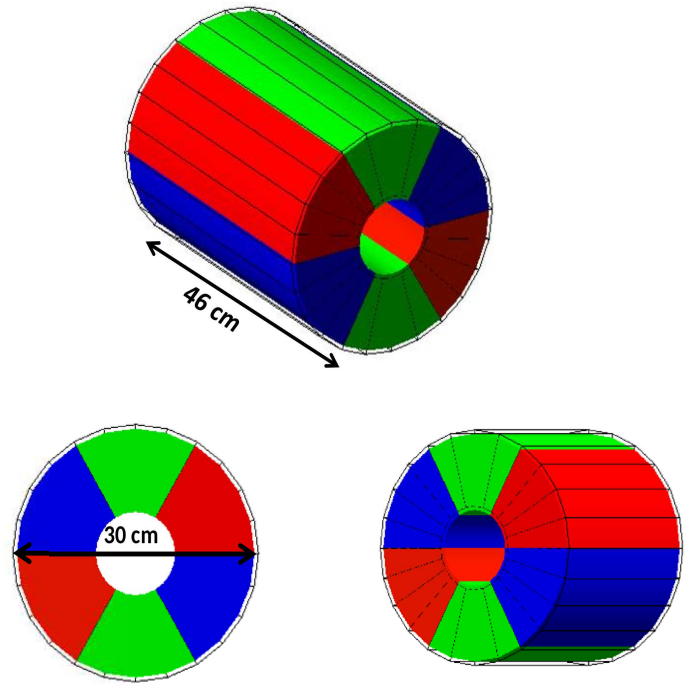}
\caption{\label{sum} \small \sl  The schematic diagram of the SUM spectrometer with six sectors of NaI(Tl) detectors shown from different
angles. The dimensions are indicated in the diagram. }
\end{center}
\end{figure}

\begin{figure}[ht]
\begin{center}
\includegraphics[width=\textwidth]{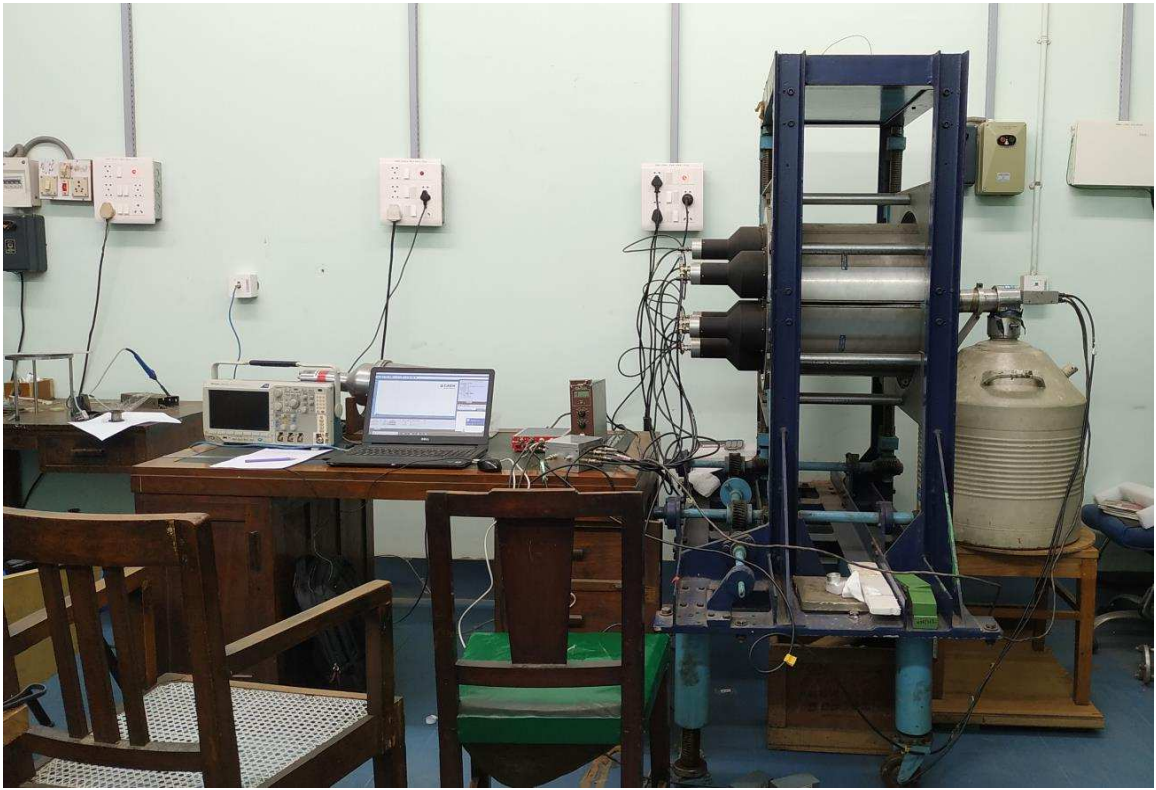}
\caption{\label{lab203} \small \sl  The shielding arrangements at Lab II with SUM spectrometer for the same 20\% relative
efficiency HPGe detector. The empty space in the SUM spectrometer borehole has been blocked with Pb blocks from one end and the HPGe detector is inserted from the other end. }
\end{center}
\end{figure}

\item{\it Setups at Lab II}
  \begin{itemize} 
\item{ No shielding:}\\
\noindent { Ambient  $\gamma$-ray radiation background have been acquired with the bare HPGe detector by placing it distant from the walls (marked as A 
in Lab II of Fig. \ref{lab}).}
\item{With passive shielding: }\\
\noindent {A NaI(Tl) cylindrical SUM spectrometer consisting of six large sectors of NaI(Tl) detectors is available at Lab II. The length of the SUM spectrometer is 46 cm, and the diameter is 30 cm (Fig. \ref{sum}). Each of the 
detector modules is coupled with one photomultiplier tube (PMT) at one end.  At the center of the cylindrical assembly, there is a borehole of 8 cm in diameter. The HPGe detector used by us has a long neck. The neck has been inserted inside the borehole. The SUM spectrometer has been utilized as passive as well as active shielding for the HPGe detector.  The experimental set up is shown in Fig. \ref{lab203}. Additional Pb bricks have been inserted in the borehole to fill the gaps to improve the passive shielding further. }

\item{With active  shielding: }\\
The SUM spectrometer is also used as an active shielding for the HPGe detector. The positive bias voltage (+1000V) of the SUM spectrometer has been provided from the DT5533EM modules developed by CAEN  for the six PMTs of the six sectors of NaI detectors. The CAEN DT5780M desktop digitizer module provided the negative bias voltage (-3000V) of the HPGe detector. The  HPGe data have been acquired, ensuring anti-coincidence with the  SUM spectrometer signals. The improvement in the background suppression
has been studied by changing the anti-coincidence time window from 1 $\mu$s to  1.5 $\mu$s, and finally to  2 $\mu$s.  The best suppression was obtained for 1 $\mu$s.  Our group has reported the preliminary results in Ref. \cite{dae2016}. 
\end{itemize}
\end{itemize}

\section{GEANT4 simulation}
 Along with the experimental measurements, to understand the effectiveness of the shielding, simulation tools have been widely used. 
In the present work, the simulations have been carried out using the GEANT4 Monte Carlo toolkit \cite{geant4}, which enables accurate simulation of 
the passage of particles through matter. Low energy electromagnetic physics models (valid from 250 eV to 100 GeV) have been used to model the 
photon interactions with the detector throughout the simulations. The simulation has been done using a series of classes like detector construction 
and building material, particle and physics process definition, particle tracking, event action, etc. The $\gamma$-ray photons are created using the 
General Particle Source (GPS) module in GEANT4. The energy deposition is recorded step-by-step and after that added for each event. The energy resolution and cut off values are incorporated properly in the simulation. Two simulation model geometries have been used to optimize the shielding. They are discussed one by one below.
\begin{figure}[ht]
\begin{center}
\includegraphics[width=11.0cm,keepaspectratio]{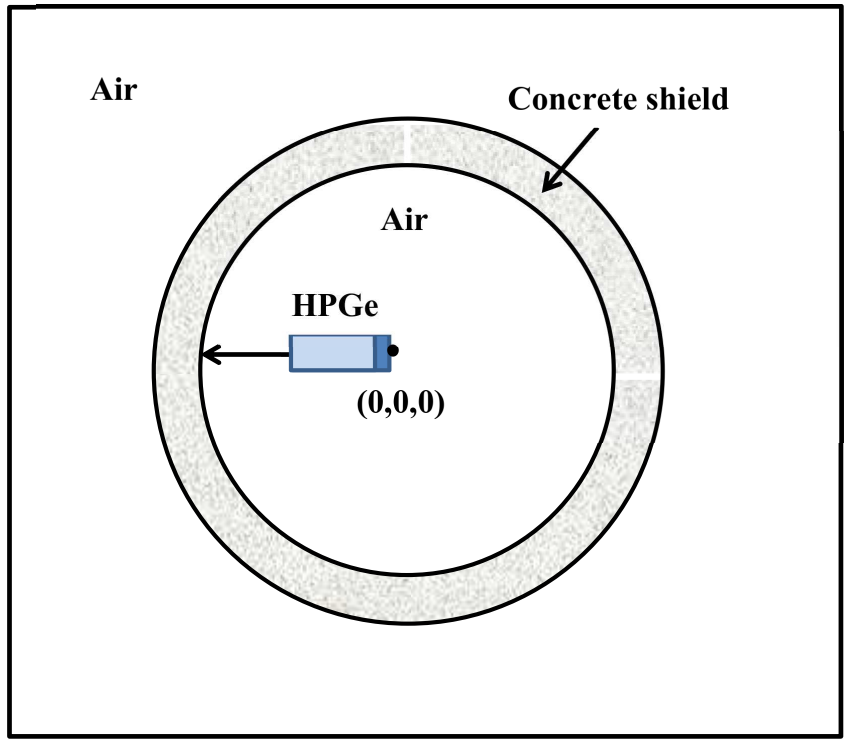}
\caption{\label{spherical_shell} \small \sl  A cross- sectional view of spherical shell geometry model showing the concrete shield.
The HPGe detector is placed at the center of the sphere. The radioactive particles are isotropically
distributed within the concrete material. The entire design is enclosed in a box, filled with air.
The figure is  drawn not to scale.
}
\end{center}
\end{figure}

\subsection{Model Geometry 1}
\begin{figure}[ht]
\begin{center}
\includegraphics[width=\textwidth]{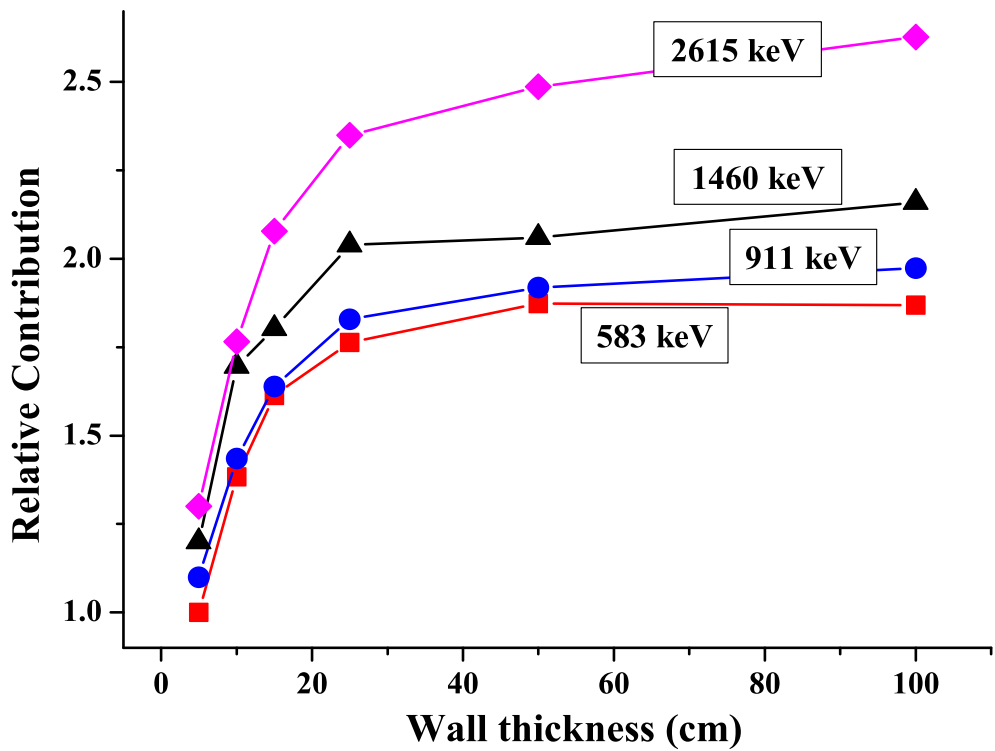}
\caption{\label{wall} \small \sl  Variation in the yield of $\gamma$-rays originating from different radionuclides (583 keV, 911 keV and 2615 keV from $^{232}$Th and 1460 keV from $^{40}$K) 
as a function of the concrete wall thickness. From the figure, it is clear that the contribution from
different $\gamma$-rays saturate after a certain thickness. The yields for 5 cm wall thickness have been considered as 1 for all the 
$\gamma$-rays. However, for better clarity of the figure, a factor of 0.1 has been added to the ratio for 911 keV, 0.2 for 1460 keV and 0.3 for 2615 keV.  }
\end{center}
\end{figure}

In the case of outdoor open fields, natural background radiations can be simulated reliably with the assumption of an infinite half-space 
source, which means a 2$\pi$- source geometry. The indoor source geometry is not easy to be realistically modeled. In Ref. \cite{Masahiro},
the authors have used an approximate 4$\pi$- geometry irradiation source for the radiation field in the buildings surrounded by concrete walls.
In the present work also, a spherical concrete layer with an isotropic distribution of natural radioactive nuclei in it has been assumed.  The HPGe detector with 20$\%$ relative efficiency is placed in the middle of the spherical shell. The spherical layer is made of concrete with a density of 2.35 g/cm$^{3}$. The composition by weight of ordinary concrete as provided in Ref. \cite{ncrp} 
is considered in the present simulation (Table \ref{concrete}). The model geometry is shown in Fig. \ref{spherical_shell}.
 The $\gamma$-ray photons have been generated using the GPS module.  The concrete wall acts as an isotropic extended $\gamma$-ray radiation source. 
Different wall thicknesses have been considered. The photon energy spectra are simulated for intense $\gamma$-rays emitted by long-lived 
radioactive nuclei, like $^{40}$K, $^{232}$Th and  $^{226}$Ra or their progenies, which are typically found in room background spectra.
The variation of photopeak areas of $\gamma$-rays of various energies are
plotted as a function of wall thickness (Fig. \ref{wall}).

\begin{table}
\centering
\caption{\label{concrete} Elemental composition of ordinary concrete \cite{ncrp} which has been considered in the present simulation.}
\vspace{0.4 cm}
\begin{tabular}{ccc}
\hline
Element   &   Amount (g in each 2.35 g) & Weight fraction\\
\hline\hline\\
Hydrogen & 0.013  &  0.005 \\
Oxygen & 1.165  &  0.496 \\
Silicon & 0.737  &  0.314 \\
Calcium & 0.194 &  0.083 \\
Sodium & 0.04  &  0.017 \\
Magnesium& 0.006  &  0.002 \\
Aluminium & 0.107 &  0.046 \\
Sulphur & 0.003 &  0.001 \\
Potassium & 0.045  &  0.019 \\
Iron & 0.029  &  0.012 \\

\hline\\
\end{tabular}
\end{table}

\begin{figure}[ht]
\begin{center}
\includegraphics[width=12.0cm,keepaspectratio]{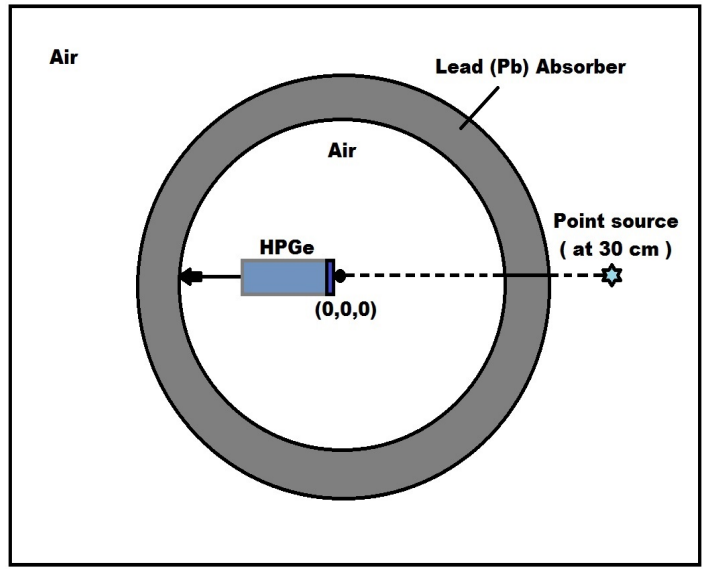}
\vspace*{-3.7cm}
\caption{\label{geometry_2} \small \sl  A cross- sectional view of spherical shell geometry model showing the Pb shield.
The HPGe detector is placed at the center of the sphere. The point radioactive source is placed at a distance of 30 cm (from the center) outside the
Pb shield. The entire design is enclosed in a box, filled with air.
The figure is  drawn not to scale.
}
\end{center}
\end{figure}
\subsection{Model Geometry 2}
\begin{figure}[ht]
\begin{center}
\includegraphics[width=14.0cm,height=10.0cm]{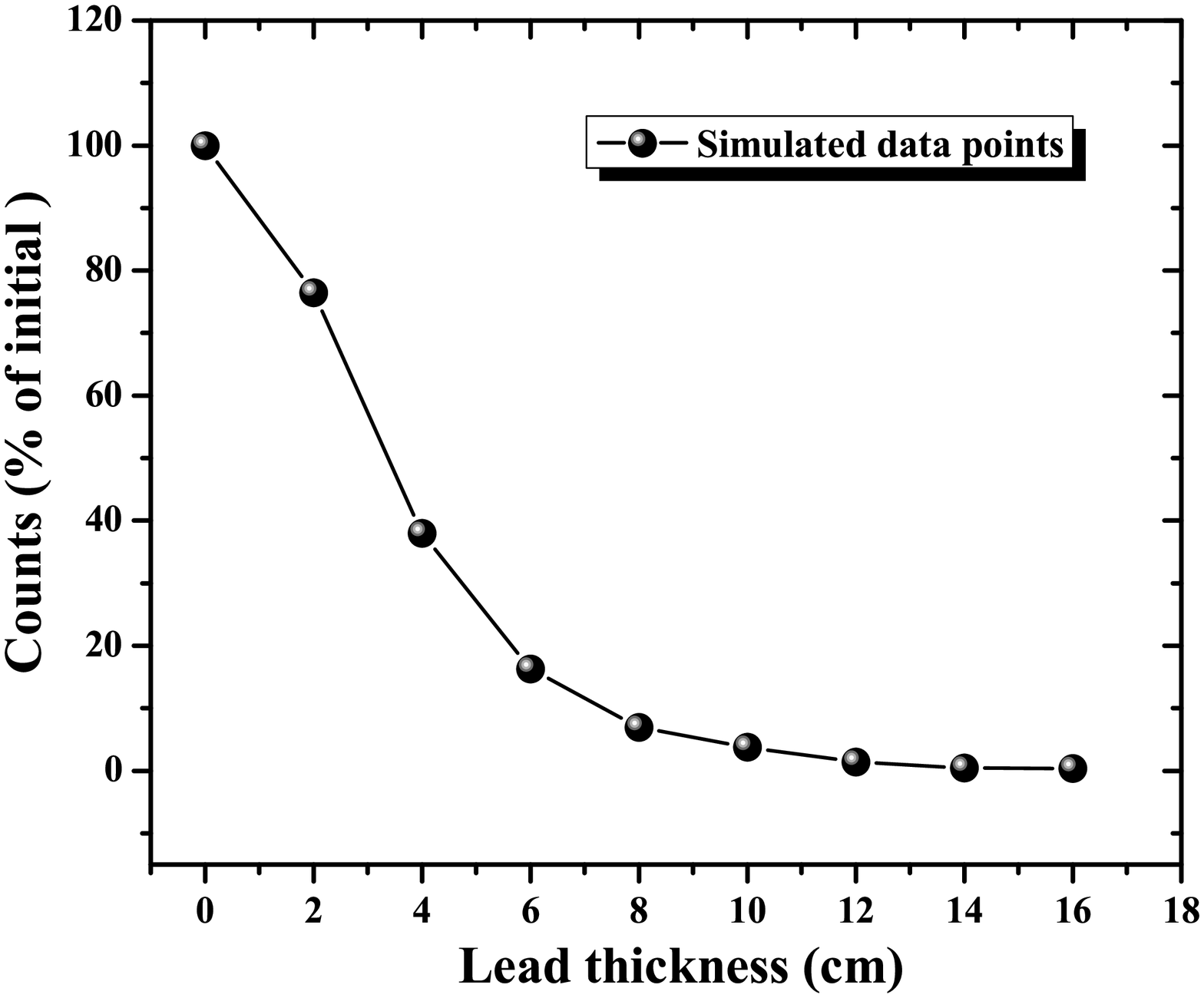}
\caption{\label{lead_shield} \small \sl The percentage of total counts detected in the detector with respect to the incident flux as
a function of lead shielding thickness. The energy spectrum in each case is simulated at a fixed
source position with $5\times 10^6$ incident $\gamma$-ray photons with energy 2.615 MeV. The number of photons reaching the detector in the absence of a Pb absorber
(0 cm of Pb thickness) has been normalized to a value of 100 $\%$ detection.  }
\end{center}
\end{figure}

In the second case, the model geometry has been built with a Pb shielding layer surrounding the  HPGe detector with 20\% relative efficiency 
(80 cm$^{3}$ volume). A point source is placed at 30 cm from the detector front face (Fig. \ref{geometry_2}). The highest energy $\gamma$-ray 
available in the room background is 2.615 MeV. So, to check the effectiveness of the lead shielding, $5\times 10^6$ $\gamma$-ray photons of 2.615 MeV energy have been thrown isotropically from the point source in 4$\pi$- direction. The total numbers of photons reaching the detector are recorded. The thickness of the lead shield is varied from 0 cm to 16 cm. Percentage of the initial count of $\gamma$-ray photons, which are finally detected by the detector is plotted in 
Fig. \ref{lead_shield}.

\begin{figure}[ht]
\begin{center}
\includegraphics[height=10.0cm, width=14.0 cm]{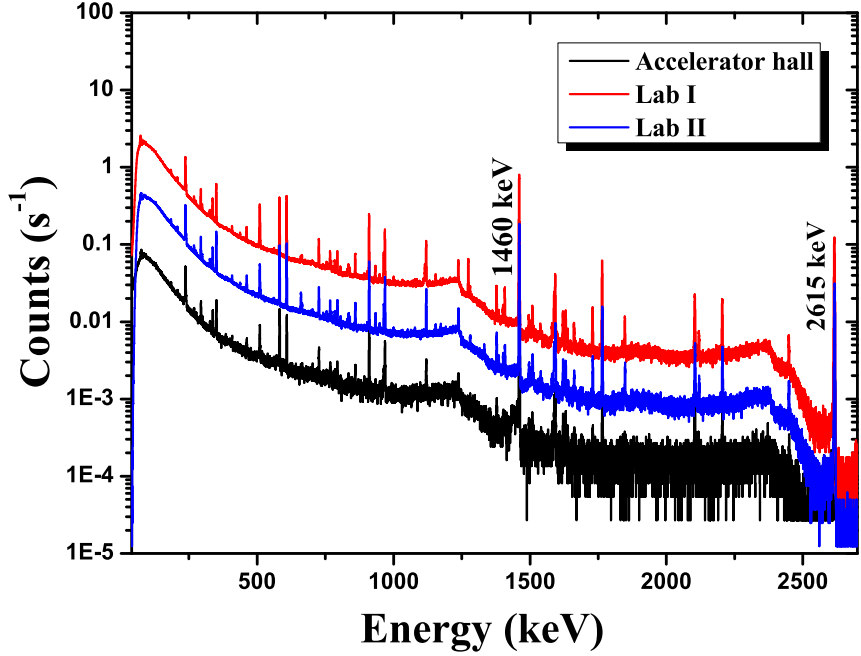}
\caption{\label{accehall_vs_lab} \small \sl  Comparison of normal $\gamma$-ray background at FRENA building accelerator hall, Lab I and Lab II respectively. The two intense photo-peaks
of  1460 keV and 2615 keV $\gamma$-rays are shown in the figure. 
}
\end{center}
\end{figure}

\begin{figure}[ht]
\begin{center}
\includegraphics[width=\textwidth]{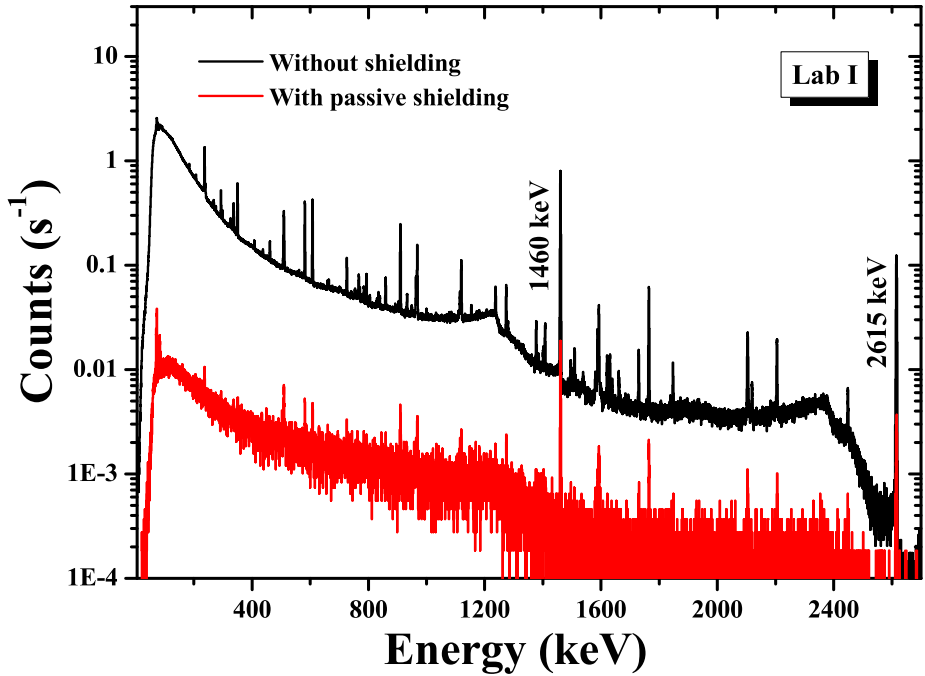}
\caption{\label{lab362_comparison} \small \sl  The spectra drawn with a black line correspond to room background without any shield at Lab I, and that with a red line correspond to data acquired with the HPGe in a Pb shielding. The overall background suppression is ten-fold in the presence of a passive Pb shield. }
\end{center}
\end{figure}

\begin{figure}[ht]
\begin{center}
\includegraphics[width=\textwidth]{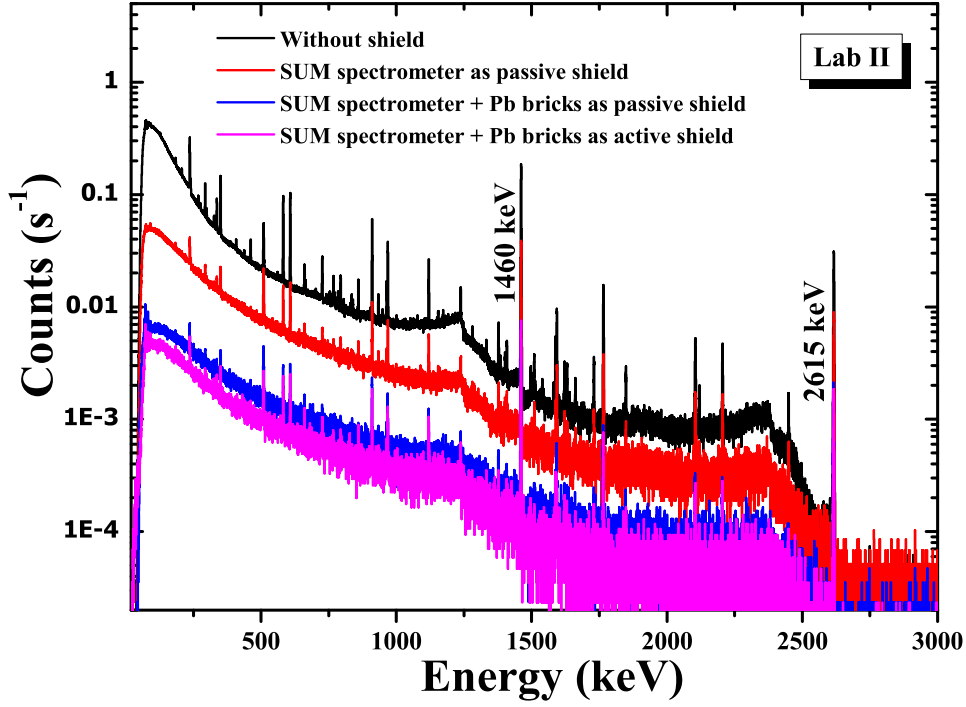}
\caption{\label{lab203_comparison} \small \sl  The spectra drawn with (i) black line corresponds to room background without any shield at Lab II, (ii) red line corresponds to  SUM spectrometer as passive shielding, (iii) blue line corresponds to  SUM spectrometer as passive shielding + additional Pb bricks and (iv) magenta line corresponds to SUM spectrometer with extra Pb bricks + active shielding with 1 $\mu$s time window. The total background suppression using this set up is 22 fold. 
}
\end{center}
\end{figure}

\begin{table}
\begin{center}
\caption{\label{cps} The count rates ($s^{-1}$) for all configurations in energy range 50-3000 keV. The configurations in detail are mentioned in the footnote below the table. }
\vspace{0.4 cm}
\begin{tabular}{cccccccc}
\hline 
Energy range &\multispan{7} \hfil Configurations \hfil \\
& C1\footnotemark[1]& C2\footnotemark[2]& C3\footnotemark[3]& C4\footnotemark[4]& C5\footnotemark[5]& C6\footnotemark[6]& C7\footnotemark[7]\\
(keV)&&&&&&\cr
\hline
50-500&1207.94&10.47&42.08&253.68&43.69&7.04&4.71  \\
500-1000&126.10&3.15&4.47&28.15&9.58&1.93&1.26    \\
1000-2000&68.77&1.95&2.49&15.54&4.95&1.00&0.69     \\
2000-3000&9.98&0.31&0.38&2.28&0.83&0.17&0.12       \\
\hline
\end{tabular}
\end{center}
\footnotemark[1]{Lab I: without any shielding } \\
\footnotemark[2]{Lab I: with nearly 7 cm of Pb shielding } \\
\footnotemark[3]{FRENA Accelerator hall (A) without shielding } \\
\footnotemark[4]{Lab II: without any shielding} \\
\footnotemark[5]{Lab II: SUM spectrometer as passive shield } \\
\footnotemark[6]{Lab II: SUM spectrometer as passive shield + Pb bricks } \\
\footnotemark[7]{Lab II: SUM spectrometer as both passive shield + active shield + Pb bricks } \\
\end{table}

\begin{figure}[h]
\begin{center}
\includegraphics[height=10.0cm, width=14.0 cm]{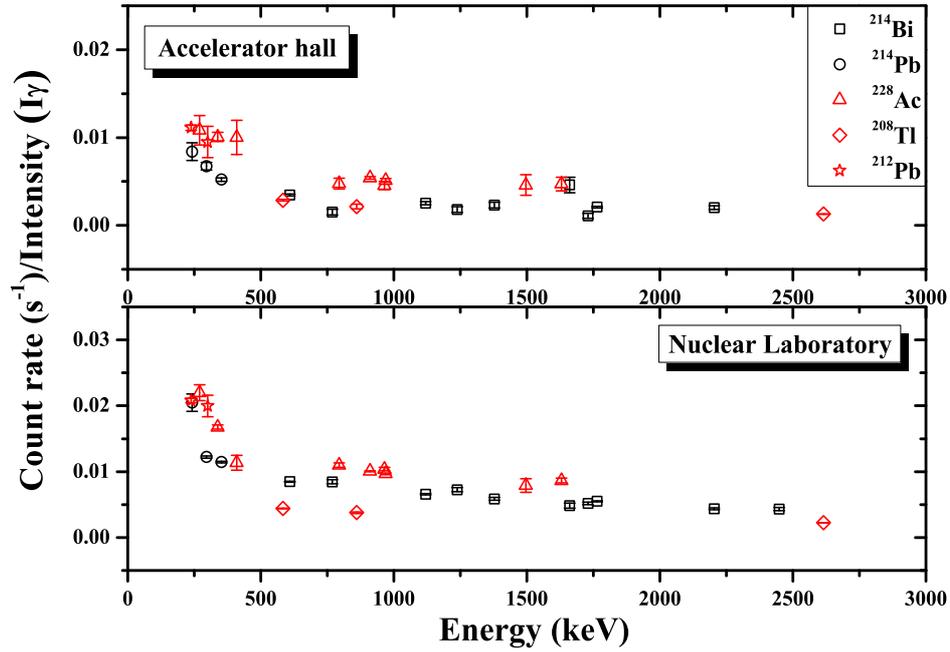}
\caption{\label{ratio_cps_vs_intensity} \small \sl  
.The ratio of the background count rates ($s^{-1}$) to the intensities of  $\gamma$-rays emitted by radioactive progenies  $^{214}$Bi, $^{214}$Pb of $^{238}$U decay series and $^{228}$Ac, $^{208}$Tl, $^{212}$Pb of $^{232}$Th decay series have been plotted as a function of energy. Data have been acquired at accelerator hall of the FRENA building and Lab I. The nature of the plots, indicates a similar distribution of the 
long-lived radionuclides in both the
indoor environments.}

\end{center}
\end{figure}

\section{Results and Discussions} 

\subsection{Experimental Results}
The indoor background spectra have been measured at different locations of the FRENA building like accelerator hall (A), beam hall middle (B), near beam hall wall (C) and control room (D) as shown in 
Fig. \ref{hall}. As expected, the photopeak count rates of relatively low energy ($<$ 1.5 MeV) $\gamma$-rays are highest in the counting room (D). However, the variation of the photopeak count rate of 
2.615 MeV $\gamma$-ray is within 4$\%$ between different measurement positions (B, C, and D). The photopeak count rate of 2.615 MeV  $\gamma$-ray  in the  accelerator hall near the concrete wall (A)
is 25$\%$ less than the other locations.  The background spectrum measured in the FRENA accelerator hall is shown in Fig. \ref{accehall_vs_lab} in comparison with the data acquired in the two 
nuclear physics laboratories.

From Fig. \ref{accehall_vs_lab}, it is clear that the count rate at FRENA accelerator hall is lower than the Lab I and II. The FRENA  hall, Lab I, and  Lab II are not of similar sizes.
 The Lab I and II are enclosed by standard red brick walls ($\approx $ 25 cm thickness), while the FRENA hall is surrounded by 1.2 m thick concrete.  
Thus the number of radioactive nuclei present in the walls is also different. Although the thick concrete walls contain more radioactivity in itself, due to larger volume, they also shield the environmental radiation generated from surroundings from reaching the detector. The brick walls, in that way, can not act as effective shield. Moreover, the background radiation also varied at FRENA accelerator hall and the two laboratories due to the larger distance of the detector from the surrounding walls in FRENA. The detection efficiency of a detector reduces sharply with the square of the distance from the source. 

The HPGe detector has been taken to the nuclear physics laboratories, 
to test the effectiveness of passive and active shieldings. The results are discussed below.

To check the shielding effect of  Pb bricks, the normal room background spectrum without any shielding at Lab I is compared with
the spectrum with the shielding (Fig. \ref{lab362}). The comparison between the two spectra is shown in Fig. \ref{lab362_comparison}. In the presence of Pb shield, the areas of the most intense peaks in the $\gamma$-ray background spectrum, 1.460 MeV and 2.615 MeV, are reduced by a factor of 45 and 36, respectively.

The passive and active shielding set up at Lab II, shown in Fig. \ref{lab203} has also been utilized for background suppression.
The total integral count rate is reduced by a factor of 10 after placing the HPGe detector inside the borehole of the SUM spectrometer. The reduction factor increases to 15 after adding some additional Pb blocks inside the borehole. Finally, we have used the SUM spectrometer both as a passive and active shield. It improves the suppression further by a factor of 22 compared to normal room background. The comparisons between the spectra at different stages are shown in Fig. \ref{lab203_comparison}. We have checked the integral count rate by varying the time window in the data acquisition system from 1 $\mu$s to 2 $\mu$s. We get the best suppression with 1 $\mu$s anti-coincidence time window.

For a detailed analysis, the spectra covering the entire range (50-3000 keV) have been divided into four parts -- a) 50-500 keV, b) 500-1000 keV, c) 1000-2000 keV, and d) 2000-3000 keV. The count rates 
($s^{-1}$) for all setups discussed above  are given in Table \ref{cps}.  

The values of the ratios of the background count rates to the intensities of decay $\gamma$-rays of $^{214}$Bi, $^{214}$Pb of $^{238}$U decay 
series and $^{228}$Ac, $^{208}$Tl, $^{212}$Pb of $^{232}$Th decay series have been plotted as a function of energy, similar to  Ref. \cite{Radulescu} for accelerator hall and Lab I. 
Both plots are shown in Fig. \ref{ratio_cps_vs_intensity}. For each location, data for $^{232}$Th and $^{238}$U  decay series follow similar shaped efficiency function. It indicates that
the long-lived $^{232}$Th and $^{238}$U  radionuclides are distributed similarly  in the FRENA hall. It is also true for  Lab I. Moreover, as the same detector has been used in both the locations,
the efficiency curves generated for FRENA accelerator hall and Lab I are also similar. The smaller distance of the walls of Lab I from the detector resulted in higher efficiency of detecting background  $\gamma$-ray radiation compared to the FRENA halls.

\begin{figure}[ht]
\begin{center}
\includegraphics[width=\textwidth]{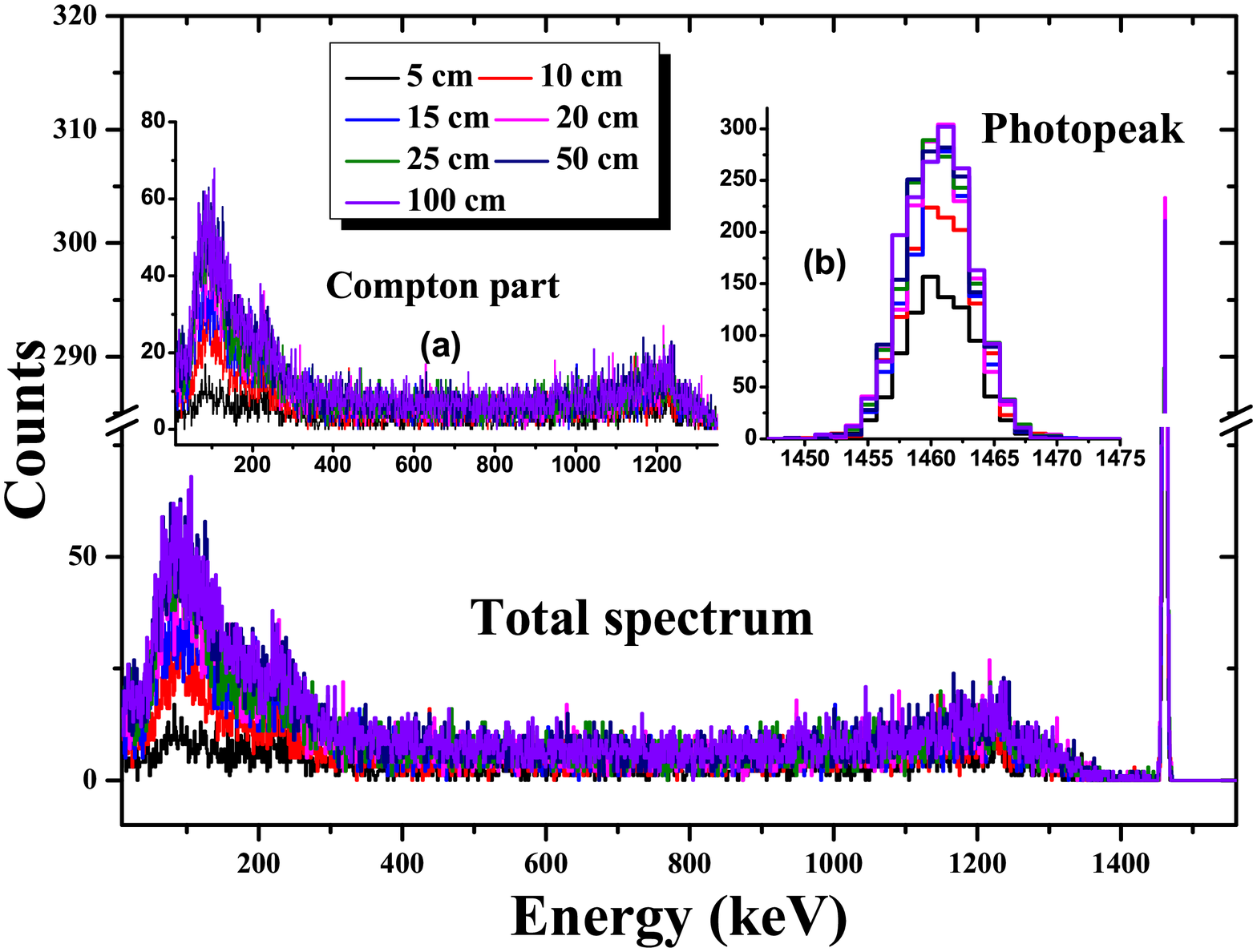}
\caption{\label{1460_variation} \small \sl  Energy distribution spectra of 1.460 MeV at different concrete wall thicknesses with
same activity concentration of $^{40}$K - (a) in case of scattered part of $\gamma$-ray radiation, it increases as the wall thickness
becomes thicker, (b) the saturation in full energy photopeak (FEP) height beyond 25 cm is seen. 
}
\end{center}
\end{figure} 

\subsection{Simulation Results}
The simulated photon energy spectra with model geometry 1 are shown in Fig. \ref{1460_variation} for various wall thicknesses, 5, 10, 15, 20, 25, 50 and 100 cm.  In all cases,  we have considered only the 1.460 MeV $\gamma$-ray from a $^{40}$K source with the same concentration per unit thickness of the wall. The Compton scattered part of the 1.460 MeV $\gamma$-ray is shown in the inset of Fig. \ref{1460_variation}(a). As the wall thickness increases, the Compton scattered part increases sufficiently; however, the FEP count saturates beyond 25 cm of wall thickness 
(Fig. \ref{1460_variation}(b)). 

In Fig. \ref{wall}, the relative contribution of $\gamma$-rays originated from different progenies of long-lived radioisotopes are plotted as a function of wall thickness. 
From Fig. \ref{wall}, it is clear that the saturation in FEP area is achieved earlier at smaller wall thicknesses for lower  FEP energy. So, $\gamma$-rays coming beyond 25 cm of the wall do not contribute to the indoor energy spectrum as they are scattered and absorbed within the wall. With the 1.2 m thick concrete wall, the $\gamma$-rays coming from outside environment till 3 MeV, are almost completely absorbed. \\

The percentage of the total initial counts of 2.615 MeV $\gamma$-ray photons, which are finally detected by the detector, is plotted in 
Fig. \ref{lead_shield} as a function of the thickness of the lead absorber.
With the geometry model 2, the initial $\gamma$-ray flux is reduced to 0.33\% after
passing through nearly 16 cm of a lead shield.

\section{Summary and Future plan}

Background spectra have been taken at different positions of the FRENA building and also at our nuclear physics laboratories. The count rate in the FRENA experimental hall is lower than in our
nuclear physics laboratory due to the differences in size and construction materials. Next, the integral count rates have been calculated for various energy ranges in different shielding configurations. The shielding arrangements which are set up at our laboratory using different accessories will be useful to decrease the indoor background further at FRENA building. The experimental data confirm that long-lived uranium and thorium are uniformly distributed in the laboratory environment.
Simulations have been done to understand and study the shielding arrangements further. The effectiveness and the contribution of the concrete shielding in the natural $\gamma$-ray background has also been studied using the 4$\pi$- geometry source irradiation. From the simulation, it is clear that the 1.2 m thick concrete shielding protects the inside from the outside environmental radiation, and only radionuclides distributed up to 25 cm inside the wall contribute to the indoor $\gamma$-ray background. The simulation is also useful to estimate the passive shielding thickness of the real experimental setup. The initial $\gamma$-ray flux is reduced to 0.33$\%$ in the presence of 16 cm thick lead absorber ( in case of a 2.615 MeV $\gamma$-ray).  \\

The background study is beneficial to plan the low energy astrophysical experiments at FRENA. In the future, we will use the shielding setups in the in-beam experiments to eliminate the background $\gamma$-ray events from the actual events.  

\section{Acknowledgments}
The authors would like to thank Ms. Sangeeta Das, Mr. Arkajyoti De, Dr. Sudatta Ray, Mr. Hitesh Rattan, Ms. Prajnaparamita Das, Mr. Samiran Malgope, 
Mr. Abu Sufiyan and Mr. Suraj Kumar Karan, for their kind help and cooperation during the experiments. We want to thank Mr. Pradip Barua for technical assistance.


\begin{thebibliography}{99}

\bibitem{Radulescu} I. Radulescu, A.M. Blebea-Apostu, R.M. Margineanu, N. Mocanu, \emph{Background radiation reduction for a high-resolution 
gamma-ray spectrometer used for environmental radioactivity measurements, } \href{https://doi.org/10.1016/j.nima.2013.03.024}{Nucl. Instr. Meth. A {\bf715} (2013) 112.}

\bibitem{Britton} R. Britton, J.L. Burnett, A.V. Davies, P.H. Regan, \emph{Monte-Carlo based background reduction and shielding optimisation for a 
large hyper-pure germanium detector,} \href{https://doi.org/10.1007/s10967-013-2572-1}{J Radioanal Nucl Chem {\bf298} (2013) 1491.}

\bibitem{caciolli} A. Caciolli et al., \emph{Ultra-sensitive in-beam $\gamma$-ray spectroscopy for nuclear astrophysics at LUNA,} \href{https://doi.org/10.1140/epja/i2008-10706-3}{Eur. Phys. J. A {\bf39} (2009) 179.}

\bibitem{Groom} D.E. Groom, N.V. Mokhov, S.I. Striganov, \emph{Muon stopping power and range tables 10 MeV--100 TeV,} \href{https://doi.org/10.1006/adnd.2001.0861}{Atomic Data and Nuclear Data Tables {\bf78} (2001) 183.}

\bibitem{dae2018} Sathi Sharma, Sangeeta Das, Arkajyoti De, Sudatta Ray, Prajnaparamita Das, Hitesh Rattan, M. Saha Sarkar, \emph{Measurement of gamma radiation background in a low energy accelerator facility,} \href{http://sympnp.org/snp2018}{Proceedings of the DAE-BRNS Symp. on Nucl. Phys. {\bf63} (2018) 1090.}

\bibitem{Masahiro} Masahiro Tsutsumi, Tetsuya Oishi, Nobuyuki Kinouchi, Ryuichi Sakamoto, Makoto Yoshida, \emph{Simulation of the Background for Gamma Detection System in the Indoor Environments of Concrete Buildings,} \href{https://doi.org/10.1080/1881.2001.9715143}{Jour. of Nucl. Sci. Tech. {\bf38} (2001) 1109.}

\bibitem{ndt} M.J. Martin and P.H. Blichert-toft, \emph{Radioactive atoms: auger-electron, $\alpha$-, $\beta$, $\gamma$-, and X-ray data,} 
\href{https://doi.org/10.1016/S0092-640X(70)80033-X} {Nucl. Data Tables A {\bf 8} (1970) 1.}

\bibitem{Vishwanath} Vishwanath P. Singh, A.M. Ali, N.M. Badiger, A.M. El-Khayatt, \emph{Monte Carlo simulations of gamma
ray shielding parameters of concretes,} \href{https://doi.org/10.1016/j.nucengdes.2013.10.008} {Nuclear Engineering and Design {\bf265} (2013) 1071.}

\bibitem{wiki} \href{https://en.wikipedia.org/wiki/Cosmic_ray}{https://en.wikipedia.org/wiki/Cosmic\_ray}

\bibitem{relative_effi} \href{https://www.ortec-online.com/-/media/ametekortec/other/overview-of-semiconductor-photon-detectors.pdf}{https://www.ortec-online.com/-/media/ametekortec/other/overview-of-semiconductor-photon-detectors.pdf}

\bibitem{CAEN} C. Tintori, \emph{ WP2081 Digital Pulse Processing in Nuclear Physics,} \href{https://www.caen.it/support-services/documentation-area/?type=manuals} {Rev. 2.1} (2011); \emph{ UM2606 DT5780 Dual Digital MCA User Manual,} \href{https://www.caen.it/products/dt5780/}{ Rev. 4.}

\bibitem{dae2016} M. Saha Sarkar, Arkabrata Gupta, Sangeeta Das, Sayantani Datta, Toshali Mitra, Indrani Ray, Yajnya Sapkota, J. Panja, Sujib Chatterjee, Ajay Mitra,  \emph{Characterization of NaI(Tl) Sum Spectrometer and its utilization, } \href{http://sympnp.org/snp2016}{Proceedings of the DAE-BRNS Symp. on Nucl. Phys. {\bf61} (2016) 1022.} 

\bibitem{geant4} S. Agostinelli et al., \emph{GEANT4-a simulation toolkit,} \href{https://doi.org/10.1016/S0168-9002(03)01368-8}{Nucl. Instr. and Meth. A {\bf506} (2003) 250. }

\bibitem{ncrp} NCRP Report no. 144, Radiation Protection for Particle Accelertor Facilities (2005).  






\end{thebibliography}
\end{document}